\documentclass[preprintnumbers,article,amsmath,amssymb,floatfix,10pt,prd,onecolumn,superscriptaddress,nofootinbib]{revtex4}
\usepackage{bm}
\usepackage{amsmath,bm}
\usepackage{amsfonts}
\usepackage{amssymb}
\usepackage{latexsym}
\usepackage[latin1]{inputenc}
\usepackage{graphicx}
\usepackage{amsmath}
\usepackage{palatino}
\usepackage{mathpazo}
\usepackage{textcomp}
\usepackage{float}
\usepackage{booktabs}
\usepackage{dcolumn}
\usepackage{hyperref}
\hypersetup{colorlinks,citecolor=blue}
\hypersetup{colorlinks=true,linkcolor=red,filecolor=magenta,    urlcolor=blue}
\usepackage{xcolor}
\usepackage{epsfig}
\usepackage{tabularx}
\usepackage{multirow}
\usepackage{mathtools}

\newcommand{\be}{\begin{equation}}
\newcommand{\ee}{\end{equation}}
\newcommand{\bea}{\begin{eqnarray}}
\newcommand{\eea}{\end{eqnarray}}
\newcommand{\eeas}{\end{eqnarray*}}
\newcommand{\beas}{\begin{eqnarray*}}

\begin{document}

\title{\bf Holographic inflation in non-static plane symmetric space-time }


\author{S. H. Shekh}\email{da\_salim@rediff.com}\affiliation{Department of Mathematics. S. P. M. Science and Gilani Arts Commerce College, Ghatanji, Dist. Yavatmal, Maharashtra-445301, India.}
\author{M. Muzammil }\email{muzammilsimaan96@gmail.com}\affiliation{Department of Mathematics,Jagadamba Mahavidyalaya, Achalpur City,  Achalpur.}
\author{R. V. Mapari}\email{r.v.mapari@gmail.com}\affiliation{Department of Mathematics, Government Vidarbha Institute of Science and Humanities Amravati, Amravati.}
\author{G. U. Khapekar}\email{ganeshkhapekar777@gmail.com}\affiliation{Department of Mathematics,Jagadamba Mahavidyalaya, Achalpur City, Achalpur.}
\author{A. Dixit}
\email{archana.dixit@gla.ac.in}
\affiliation{Centre for Cosmology, Astrophysics and Space Science (CCASS), GLA University, Mathura-281406,U.P., India.}
\begin{abstract}
\textbf{Abstract:}
The current analysis uses the non-static plane symmetric space-time to dynamically examine the holographic dark energy model as a candidates of IR cut-offs (specifically Hubble's and Granda-Oliveros cut-off).  Using the Markov Chain Monte Carlo (MCMC) method, we estimate the best fit values for the model parameters imposed from the combined datasets of $CC+SC+BAO$. Now, it has been found that the characteristics of space-time that have been addressed and formulated using both models are flat universe and observed that the model appears to be in good agreement with the observations.  In addition, we investigate the behavior of  equation of state  parameters along with the energy conditions. Finally, we found that in both the cut-offs the models predict that the present and late universe are accelerating and the equation of state parameter behaves like the quintessence model.

\end{abstract}


\maketitle

\section{Introduction}\label{1}
The universe is expanding at an ever-increasing rate, according to recent astrophysical observations like supernova \cite{1}, \cite{2}, the Wilkinson microwave anisotropy probe WMAP \cite{3}, large scale structure (LSS) \cite{4}, \cite{5}, fluctuations of cosmic microwave background radiation (CMBR),the slogan digital sky survey (SDSS) \cite{6} and the Chandra X- ray observatory \cite{7}. In most cases, there are two approaches that can be taken to explain the universe's behavior: One is the mysterious negative pressure energy component known as Dark Energy (DE), which is important because Einstein's field equations change the energy momentum tensor. However, the underlying mechanism of the accelerated expansion is still up for debate. Numerous models have been developed to explain the current universe acceleration. Modified gravity, or MG, is the second strategy, and it involves altering the geometry of space-time as described in Einstein's equations.
DE exerts a significant negative force, resulting in an anti-gravity effect that drives the acceleration \cite{8}, \cite{9}, \cite{10}, despite the fact that matter is attracted by gravity. Cosmologists have argued that the cosmological constant ($\Lambda$) is the most suitable candidate for the DE based on various observational findings. This argument is based on the fact that the negative pressure and constant energy density during cosmic evolution. However, this DE idea can be used, and if history is checked, it frequently exhibits "fine tuning" and "cosmic coincidence" puzzles \cite{11}. As a result, a number of DE models can be distinguished by the Equation of State (EOS) parameter $\omega = \frac{p}{\rho}$. Astrophysical data indicate that this parameter is very close to -1 and that the constant energy density with negative pressure in the case $\omega = -1$, which is the cosmological constant, occurs during cosmic evolution. In addition to taking into account the cosmological constant with $\omega = -1$ ($\Lambda$CDM model), dynamical DE models fall into distinct categories:

\begin{itemize}
	\item The scalar fields including Phantom with $\omega <-1$,  Quintessence with $-1<\omega < -\frac{1}{3}$, K-essence, Tachyon, Quintom,, Dilaton" and so forth \cite{12,13,14,15,16,16a}.
	\item Interacting models of DE such as the  Braneworld models, Chaplygin gas models, Holographic and Agegraphic model \cite{17,18,18a,18b,19}.
	\item Some alternate limits" $-1.67<\omega<-0.62$ and $-1.33< \omega<-0.79$" which are obtained from observational results given \cite{20, 21}.
\end{itemize}
Thus, based on theoretical arguments and experimental evidence, the existence of an anisotropic universe with anisotropic pressure that eventually approaches an isotropic universe has been established. As a result, universe models with an anisotropic background in the presence of DE merit investigation. The well-known holographic principle (HP) was first proposed by Gerard't Hooft, who was influenced by studies on the thermodynamics of black holes \cite{22, 23}. The holographic principle, which corresponds to a theory locating on the boundary of that space, states that all of the information contained in a volume of space can be represented as a hologram. This is a modern version of "Plato's cave." Cohen and others \cite{24} have argued that the DE ought to adhere to the holographic principle and be restricted by the infrared cut-off. The Holographic Dark Energy (HDE) model was proposed by Li \cite{25} following the application of the holographic principle to the DE issue. The reduced Planck mass and the cosmological length scale, which is chosen as the universe's future event horizon, are the only two physical quantities on the boundary of the universe in this model that affect DE's energy density \cite{26}. The first theoretical DE model based on the holographic principle is the HDE model and is concurrently in good agreement with the current cosmological observations. Because of this, HDE is a very strong DE contender. The HDE paradigm has received a lot of attention and been studied extensively in recent years. In any case, Pavon and Zimdahl \cite{27} have then shown that in the connecting HDE model, the recognizable proof of cosmological length scale with the Hubble skyline can likewise drive speeding up universe. Later, Karwan \cite{28} investigated the interacting HDE model, using the IR cut-off as the Hubble horizon, and discovered the fixed points as well as the conditions under which they maintain stability. He demonstrated that cosmic evolution can reach close to the present epoch and the coincidence problem can be alleviated for a small range of parameters in the interacting model. As a result, the parameters, not the initial conditions, caused the coincidence problem. Shekh and Chirde \cite{29}. Wang et al. \cite{30} , Iqbal and Jawad \cite{31}, Sharma and Dubey \cite{32},  Moradpour et al. \cite{33}. Some authors are Bhattacharjee \cite{34} who have looked at various holographic dark energies by using Hubble's and Granda-Oliver's infrared cutoffs, respectively, to compare and contrast.

\section{Cosmology with non-static plane symmetric space-time}\label{2}
Consider a non-static plane symmetric  Riemannian space-time described by the line element \cite{34a, 34b}
\begin{equation}\label{e1}
ds^2=e^{2h}\left(dt^2-dr^2-r^2d\theta^2-s^2dz^2\right)
\end{equation}
where $r,\theta,z$ are the usual cylindrical polar coordinates, $h$ and $s$ are the functions of cosmic time $t$ alone.\\
The Einstein's field equation is given by 
\begin{equation}\label{e2}
R_{ij}-\frac{1}{2}R g_{ij}=-\left( T_{ij}+\bar{T}_{ij}\right)
\end{equation}
where $R_{ij}$ is the Ricci tensor, $R$ is the Ricci scalar and $ T_{ij}$ and $\bar{T}_{ij}$ are the energy momentum tensor of dark matter and HDE respectively.\\
The energy momentum tensor for dark matter and HDE are defined as
\begin{equation}\label{e3}
T_{ij}= \rho_{m} u_{i} u_{j} {\;\;\;\;\;\;} \text{and} {\;\;\;\;\;\;} \bar{T}_{ij}=\left(\rho_{\Lambda}+p_{\Lambda}\right) u_{i} u_{j}+p_{\Lambda} g_{ij}
\end{equation}
where $\rho_{m}$, $\rho_{\Lambda}$ are the energy densities of dark matter and HDE respectively and $p_{\Lambda}$ is the pressure of HDE, $u_{i}$ is the four velocity of the fluid. Also, $u_{i}=\delta^{i}_{4}$ is a four-velocity vector which satisfies
\begin{equation}\label{e4}
g_{ij}=u^{i} u_{j}= -x^{i} x_{j} = -1 {\;\;\;\;\;\;} \text{and} {\;\;\;\;\;\;} u^{i} x_{j}=0
\end{equation}
Sarkar et al. \cite{35} worked on  Bianchi type-I universe which is spatially homogeneous and anisotropic and containing two minimally interacting fluids, matter and HDE components, in light of the energy momentum tensor for dark matter and HDE". A correspondence between the HDE models and the generalised Chaplygin gas DE model was also established using an exact solution to Einstein's field equations. It was also discovered that the field equations' solution predicts a future singularity of the big rip type, which is consistent with the findings of recent observations. With increasing time, the coincidence parameter rises and the potential and dynamics of the scalar field, which characterises the Chaplygin cosmology, are also reconstructed. The anisotropy parameter of the cosmos approaches 0 over very long cosmic times. An interacting Holographic Polytropic gas model of DE with a hybrid expansion condition has been studied by Rahman and Ansari \cite{36} in spatially homogeneous and anisotropic Bianchi type-VI$_0$ space-time. By taking into account a hybrid expansion law that depicts a transition of the universe from its previous decelerating phase to the current accelerating phase, it was also established that the HDE and polytropic gas model of DE correspond, and that the coincidence parameter is found to be increasing with time also it has been determined that the universe's physical and geometrical characteristics are in agreement with current discoveries. 
The capacity that we have to reconstruct the potential and dynamics for the polytropic gas's scalar field, which depicts the universe's accelerating expansion, is made possible by this agreement.
With the help of equations (\ref{e3}), the components of Einstein's field equation for non-static plane symmetric  Riemannian space-time (\ref{e1}) is given by
\begin{equation}\label{e5}
e^{-2h}\left(2\ddot{h}+\dot{h}^2+2\frac{\dot{h}\dot{s}}{s}+\frac{\ddot{s}}{s}\right)=-p_{\Lambda},
\end{equation}
\begin{equation}\label{e6}
e^{-2h}\left(2\ddot{h}+\dot{h}^2\right)=-p_{\Lambda},
\end{equation}
\begin{equation}\label{e7}
e^{-2h}\left(2\frac{\dot{h}\dot{s}}{s}+3\dot{h}^2\right)=\rho_{m}+\rho_{\Lambda},
\end{equation}
where the overhead dot denotes differentiation with respect to cosmic time $t$ which measure in Gyr.\\
The total energy density fulfils the continuity equation for a universe where dark matter and DE interact as follows.
\begin{equation}\label{e8}
\dot{\rho}_{m}+\dot{\rho}_{\Lambda}+3H(\rho_{m}+\rho_{\Lambda}+p_{\Lambda})=0.
\end{equation}
We think about how dark matter and DE interact. Dark matter and DE thus do not conserve their respective energy densities individually. The expression for the constancy of matter is
\begin{equation}\label{e9}
\dot{\rho}_{m}+3H\rho_{m}=Q,
\end{equation}
\begin{equation}\label{e10}
\dot{\rho}_{\Lambda}+3H(\rho_{\Lambda}+p_{\Lambda})=-Q,
\end{equation}
where "$Q$" denotes the way dark matter and dark energy interact. Generally speaking, $Q$ should be a function with units that are the inverse of time. We select $Q=3 \delta H \rho_m$ for the sake of convenience, where $\delta$ is the coupling constant. The equation of continuity becomes the non-interacting case when $\delta=0$. Many authors \cite{37,38,39} has examined these interactions. The energy density of pressureless dark matter is calculated as based on the above interaction from the equation (\ref{e9}).
\begin{equation}\label{e11}
\rho_{m}= c_{2}a^{3(\delta-1)}
\end{equation}
From equations (\ref{e5}) and (\ref{e6}), we got the integration-based relationship between $s$ and $e^{2h}$
\begin{equation}\label{e12}
s=c \int e^{-2h}dt+c_1
\end{equation}
where $c$ and $c_{1}$ both are the constant parameters, without loss of generality we take $c_1=0$. Using the above equation (\ref{e12}), the scale factor $a=V^{1/3}$ is obtained as
\begin{equation}\label{e13}
a=e^{\frac{4}{3}h}\left(c \int e^{-2h}dt\right)^{\frac{1}{3}}
\end{equation}
The dimentionless quantity called Hubble's parameter ($H$) which related with scale factor $\left(H=\frac{\dot{a}}{a}\right)$ is obtained as
\begin{equation}\label{e14}
H=\left(\frac{2-n}{3}\right)\frac{1}{t}
\end{equation}
Applying the transformation with $a=(1+z)^{-1}$, the above expression of $H$ in $z$ can be observed as
\begin{equation}\label{e15}
H=H_0 \left(1+z\right)^{\frac{3}{2-n}}
\end{equation}
Using the above expression of Hubbles Parameter $H$, from the Eq. (\ref{e5}) the pressure of the model is observed as,
\begin{equation}\label{e16}
p_{\Lambda}=\alpha_{1}\left(\frac{1}{1+z}\right)^{\frac{3(n-3)}{2-n}}
\end{equation}
where $\alpha_{1}=\frac{3^{3-n}n(1-n)(n+3)(2-n)^{n-3}}{4c_{1}H_{0}^{n-3}}$.\\
A quantity called the equation of state (EoS) parameter is used in cosmology to explain the characteristics of dark energy, a fictitious type of energy supposed to be the source of the universe's accelerating expansion. The behaviour of DE as a function of time is described in this context by the EoS parameter. The symbol $\omega$ or $\omega_{\Lambda}$ is frequently used to denote it and defined as 
\begin{equation}\label{e17}
\omega_{\Lambda}=\frac{p_{\Lambda}}{\rho_{\Lambda}}.
\end{equation}

To investigate the cosmic parameters and the time evolution of the universe. Mathematically speaking, we also require more constraints in order to fully solve the system. The model-independent way to examine the dynamics of the dark energy model is the most well-known physical justification for adopting these constraints in literature. The fundamental idea behind this approach is to take into account a parametrization of any cosmological parameters, like the Hubble parameter. To describe challenges with cosmological investigations, such as the initial singularity problem, the challenge with all-time decelerating expansion, the horizon problem, the Hubble tension, etc., these schemes have been extensively discussed in the literature. One might refer to \cite{40,41,42,42a,42b} for a thorough overview of the various cosmological parameterization systems. Hence, in this following subsection, we will employ data sets from various cosmological measurements, or data that specifies the distance-redshift connection, to constrain the free parameters of $H(z)$.

\subsection{ Data description and Results}\label{Observation}
We have described the methodology used to calculate the Hubble's parameter in the section above. To get the best fit values, we constrained the parameters listed in $H$ in the following subsections using Cosmic Chronometric (CC) data, Standard Candles SN Ia (SC) from Pantheon, and Baryon Acoustic Oscillations (BAO)\cite{43}. In the sections that follow, we go into considerable detail about the data we used and the approach we applied.
\subsubsection{CC data}\label{CC}
In general, there are two well-established methods to estimate the value of $ H(z) $ at a particular redshift, namely BAO and differential age approach \cite{44}. We use the 31 measurements of the CC dataset in the range of redshift provided as $ 0.07<z<2.41 $. The Chi-square function is defined as follows.
\begin{equation}\label{e18}
\chi_{CC}^2=\sum_{i=1}^{31} \frac{\left[H_{t h}-H_{o b
s}\left(z_i\right)\right]^2}{\sigma_{H\left(z_i\right)}^2},
\end{equation}
where $ H_{th},H_{obs} $ and $\sigma_{H(z_{i}}  $ represent
theoretical value, observed value and stranded error respectively
of $ H $ . For the table of 31 points of $ H(z)$ data

\begin{figure}[H]
	\begin{center}
		\includegraphics[width=14cm,height=9cm, angle=0]{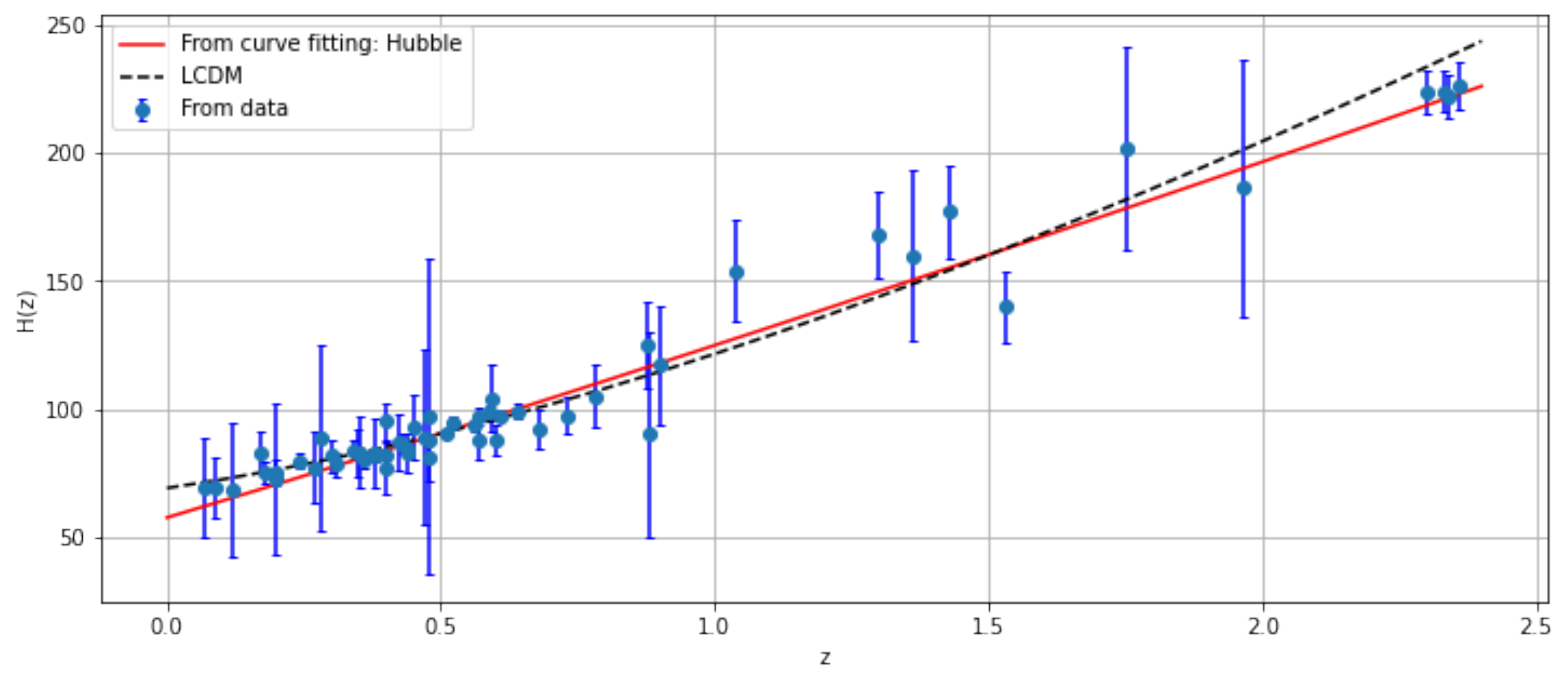}
		
		\caption{ The error
			bar for the Hubble parameter with the standard $\Lambda$CDM model are  shown in figure with combined CC+SC+BAO datasets.}
		
	\end{center}
\end{figure}

\subsubsection{SC data}
As Standard Candles (SC) we use measurements of
the Type Ia Supernovae (SNeIa) \cite{45,46,47,48,49,50}consist of 1048 measurements. They were collected from five different sub-samples PS1, SDSS, SNLS, low- $z$, and HST lying in the redshift range $0.01<z<2.3$ . The model parameters can be fitted by comparing the observed $\mu_i^{o b s}$ with theoretical $\mu_i^{\text {th }}$ value of the distance moduli

\begin{equation}\label{e19}
\mu=m-M=5 \log _{10}\left(D_L\right)+\mu_0,
\end{equation}

where $M$ and $m$ represents the absolute and apparent magnitudes respectively and $\mu_0=5 \log \left(H_0^{-1} / M p c\right)+25$ is the nuisance parameter that has been depreciate. The luminosity distance is defined by

\begin{equation}\label{e20}
D_L(z)=\frac{c}{H_0}(1+z) \int_0^z \frac{d z^*}{E\left(z^*\right)}
\end{equation}

The $\chi^{2}$ function of the (SNeIa) measurements is given as

\begin{equation}\label{e21}
\chi_{\mathrm{SN}}^2\left(\phi_{\mathrm{s}}^\nu\right)=\mu_{\mathrm{s}} \mathbf{C}_{\mathrm{s}, \text { cov }}^{-1} \mu_{\mathrm{s}}^T,
\end{equation}

where "$\mu_{\mathrm{s}}=\left\{\mu_1-\mu_{\mathrm{th}}\left(z_1, \phi^\nu\right), \ldots, \mu_N-\mu_{\mathrm{th}}\left(z_N, \phi^\nu\right)\right\}$ and the subscript 's' denotes SnIa. Keep in mind for SnIa dataset the covariance matrix is not diagonal". The distance modulus $\mu_i$ is given as $\mu_i=\mu_{B, i}-\mathcal{M}$, where  $\mu_{B, i}$ represents the maximum apparent magnitude in the cosmic rest frame for redshift $z_i$ and $\mathcal{M}$ represents the universal free parameter, by various observational ambiguity both $\mathcal{M}$ and $h$ parameters are naturally degraded with respect to Pantheon dataset, Hence one can't extract proper information about  $H_0$ from SnIa data alone.

\begin{figure}[H]
	\begin{center}
		\includegraphics[width=14cm,height=8cm, angle=0]{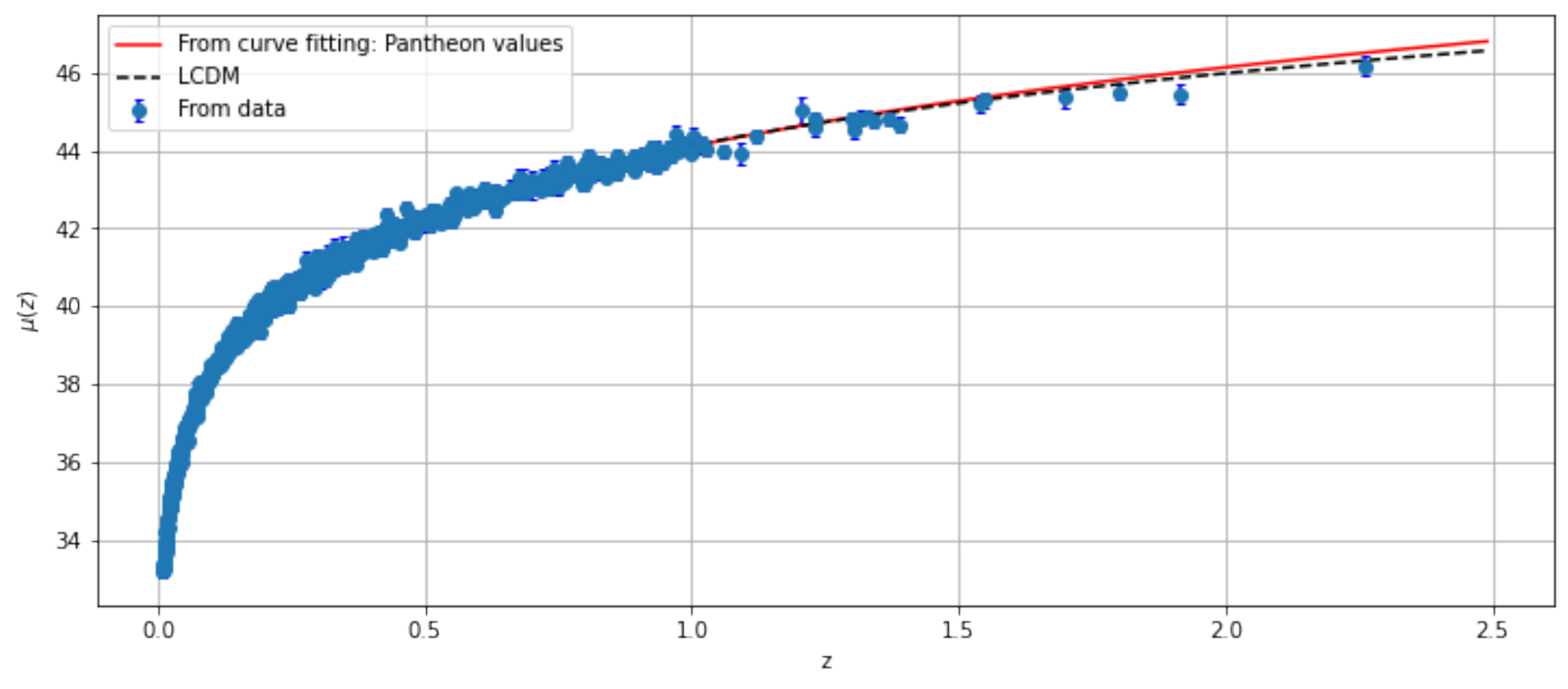}
		
		\caption{ The distance modulus versus redshift z  with the
			constraint values of combined CC+SC+BAO datasets.}
		
	\end{center}
\end{figure} 

\subsubsection{BAO data}
BAO is the direct repercussion of strong coupling between baryons and photons in the pre-recombination epoch. After decoupling of photons, the overdensities in the baryon fluid evolved and attracted more matter, leaving an imprint in the two-point correlation function of matter fluctuations with a characteristic scale of around $r_d \approx 147 \mathrm{Mpc}$ that can be used as a standard ruler and to constrain cosmological models \cite{52}. Further Studies of BAO feature in the transverse direction provide a measurement of $D_H(z) / r_d=c / H(z) r_d$, with the comoving angular diameter distance in 

\begin{equation}\label{e22}
D_M=\int_0^z \frac{c d z^{\prime}}{H\left(z^{\prime}\right)}
\end{equation}

BAO peak coordinates can be obtained by combining the angular diameter distance $D_A=D_M /(1+z)$ and $D_V(z) / r_d$ \cite{ref53}

\begin{equation}\label{e23}
D_V(z) \equiv\left[z D_H(z) D_M^2(z)\right]^{1 / 3} .
\end{equation}

We are going to use 17 uncorrelated BAO measurements, collected in, in the redshift range of $0.106<z<2.34$. Since this proves the uncorrelation of this dataset,

The $\chi^{2}$ function of the BAO measurements given as

\begin{equation}\label{e24}
\chi_{B A O}^2=\sum_{i=1}^{17}\left(\frac{D_{obs}-D_{\text {th}}\left(z_i\right)}{\Delta D_i}\right)^2,
\end{equation}

where $D_{obs}$ is the observed distance module rates at redshift $z_i(i=1, \ldots, n)$ and $D_{\text {th }}$ is the theoretical distance module from the model. 
\subsubsection{Monte Carlo Markov Chain (MCMC)}
We will apply the constraints through MCMC to evaluate our model with the data that is currently available, by minimizing the overall $chi^2$ function of the combination of Cosmic Chronometric (CC), Standard Candles (SC), and BAO defined as, 
\begin{equation}\label{e25}
\chi^2=\chi_{CC}^2+\chi_{SC}^2+\chi_{BAO}^2
\end{equation}

\begin{figure}[H]
	\begin{center}
		\includegraphics[width=13cm,height=12cm, angle=0]{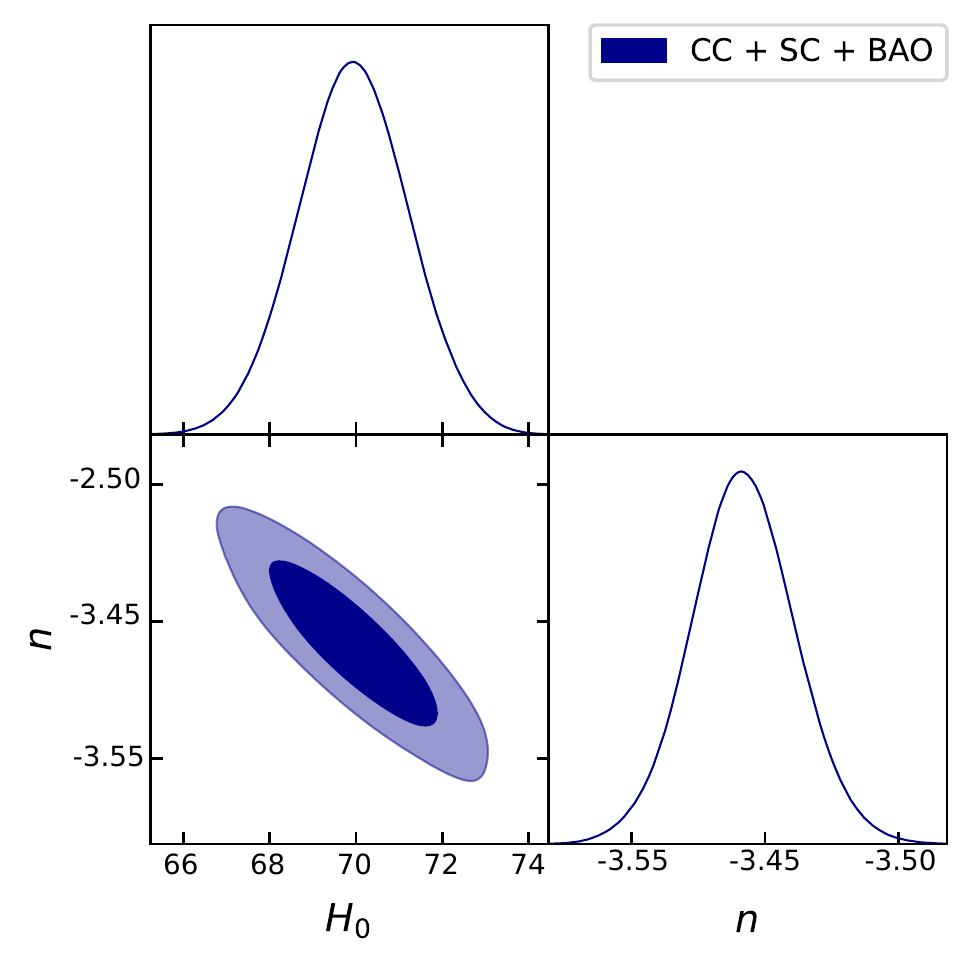}
		
		\caption{ Contour plot with 1 -$\sigma$ and 2-$\sigma$ errors for the parameters $H_{0}$,and $q$ along with the constraint values for combined CC+SC+BAO datasets. }
		
	\end{center}
\end{figure}
 As a result, we determine we determine the best fit values of the model  parameters are as $H_{0}=70.7$ and $n=-3.036$.
after applying the MCMC procedure effectively. The confidence contours and minimized posterior distribution of various cosmological parameters from the combined $CC+SC+BAO$ data sets are shown in Fig. 1. It should be emphasised that the Hubble constant, $H_0 = 100h$, is regarded to have an exciting aspect. Our final finding for $H_0$ closely reflects the value of the Planck $\Lambda$CDM estimation \cite{43} when compared to the values obtained and the one anticipated by Planck.
\section{Inflation with IR cut-offs}
The holographic dark energy model, among other dynamical dark energy models, has recently emerged as a useful method for thinking about the dark energy puzzle. Because black holes have been developed in quantum field theory, the bound on the vacuum energy of a framework with size $L$ should not exceed the restriction of the black hole mass of a similar size according to the holographic principle. The holographic principle will place a limit on the entropy of the universe in a cosmological context \cite{44}. For an effective quantum field theory in a box of size $L$ with a short distance cut-off (UV cut-off: $\Lambda$), it seems reasonable to require that the total entropy satisfy the relation.
\begin{equation}\label{eq26} 
L^{3} \Lambda^{3} \le S_{BH} = \pi L^{2} M^{2}_{p}
\end{equation}
where $S BH$ is the entropy of a black hole with a radius of $L$ that serves as a long distance cut-off (IR cut-off: $L$) and $M_p$ is the reduced Planck mass.
The zero point energy density is therefore limited by this UV-IR relationship's upper limitation as\\
\begin{equation}\label{eq27} 
\rho_{\Lambda} \le L^{-2} M^{2}_{p}
\end{equation}
The largest $L$ is chosen by saturating the bound in equation (\ref{eq27}). The holographic dark energy density is obtained as \cite{45}
\begin{equation}\label{eq28} 
\rho_{\Lambda} =\frac{3c^{2} M^{2}_{p}}{L^{2}}
\end{equation}
where $c$ is a free dimensionless $O(1)$ parameter say arbitrary parameter, the coefficient 3 is chosen for convenience and for the remaining part of the paper we will consider $c^{2}= M^{2}_{p}=1$.\\
There has been a lot of intriguing work on holographic dark energy recently, including, in \cite{46} the authors  have created a cosmological model of the universe with a constant deceleration parameter that corresponds to a purely accelerating model. Shekh et al. examined the physical acceptability of a spatially homogeneous, isotropic line element filled with two fluids and found that the region in which the stable model is presented depends on the real parameter $\delta$, for all $\delta \ge 4.5$, the models range from the $\Lambda$CDM era to the quintessence era \cite{47}.

By assuming that the deceleration parameter varies with cosmic time, Koussour et al. \cite{48} examined a holographic dark energy model with a homogeneous and anisotropic Bianchi type-I Universe and compared it to the model by examining the Jerk parameter. However there are also other remarkable studies on holographic dark energy is in \cite{49,50}.

\subsection{Hubble's cut-off}
\begin{figure}[H]
   \begin{minipage}{0.49\textwidth}
     \centering
   \includegraphics[scale=0.8]{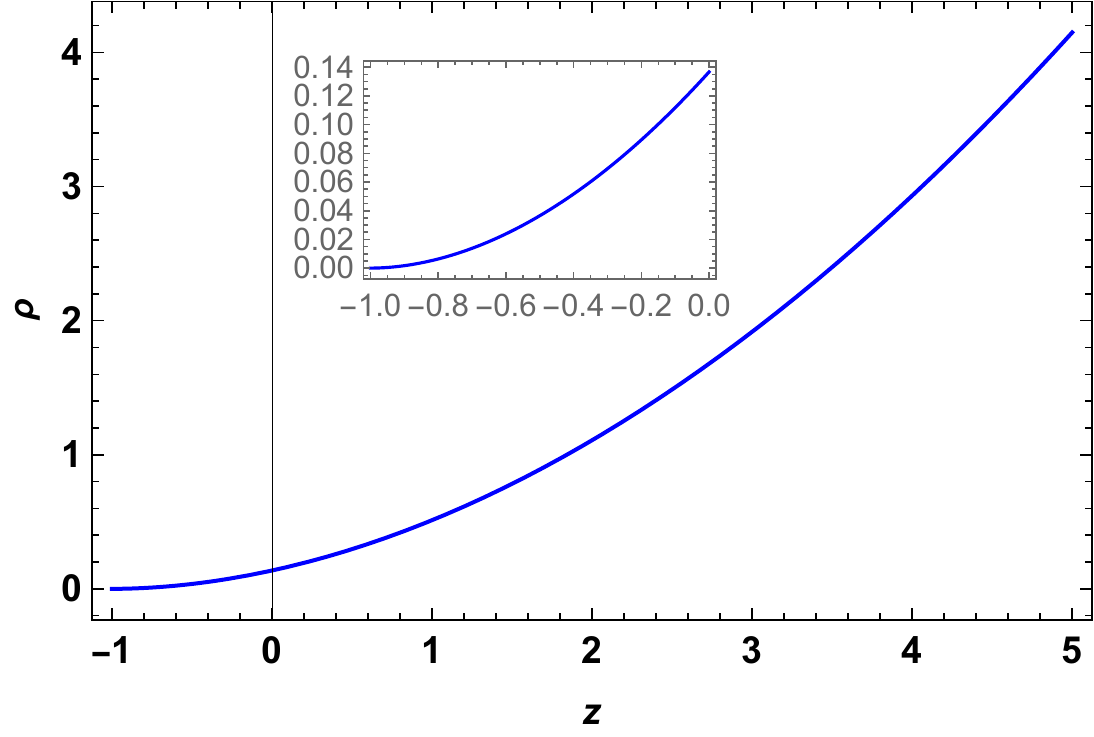}
\caption{The figure shows the evolution of $\rho(z)$ in Hubble's horizon as a potential IR Cut-off.}\label{HD}
   \end{minipage}\hfill
   \begin{minipage}{0.49\textwidth}
     \centering
 \includegraphics[scale=0.84]{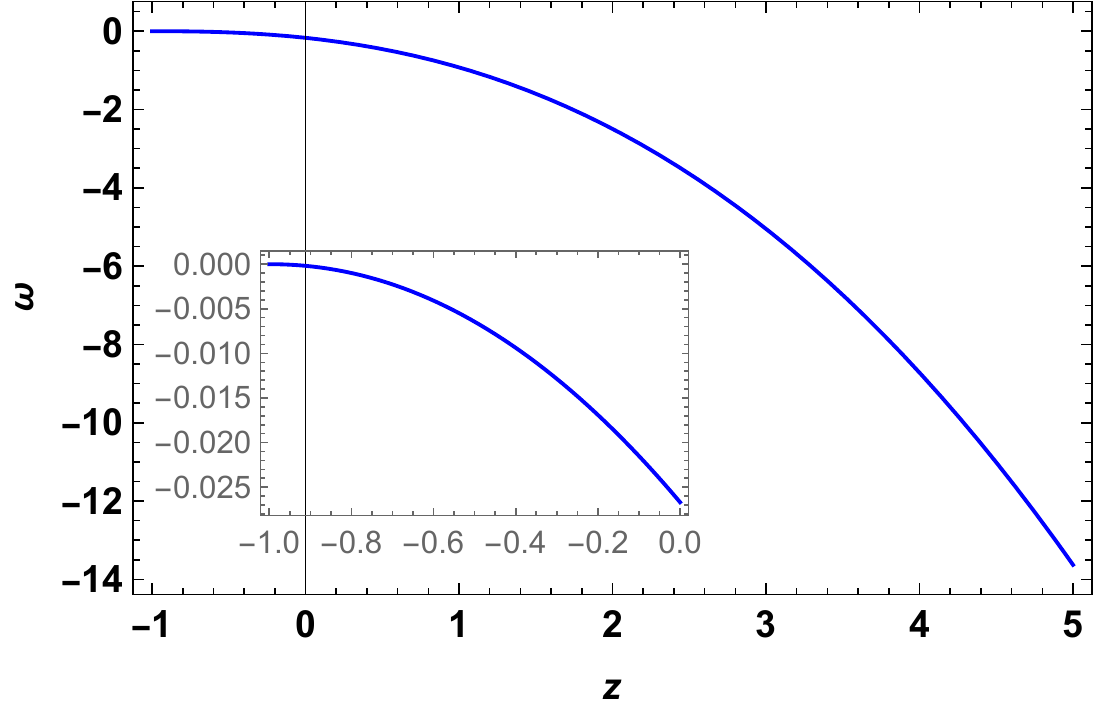}
\caption{The figure shows the evolution of $\omega(z)$ in Hubble's horizon as a potential IR Cut-off.}\label{w}
   \end{minipage}
\end{figure}

Here, some cosmological parameter are determined by using Hubble's horizon ($L=H^{-1}$) as a potential IR-cut-off.\\
In view of Eq. (\ref{e15}),  the expression of holographic dark energy density from Eq. (\ref{e29}) is obtained as,
\begin{equation}\label{e29}
\rho_{\Lambda}=3H_0^2 \left(1+z\right)^{\frac{6}{2-n}}
\end{equation}
EoS parameter,
\begin{equation}\label{e30}
\omega_{\Lambda}=\alpha_{2}\left(\frac{1}{1+z}\right)^{\frac{3(n-1)}{2-n}}
\end{equation}
where $\alpha_{2}=\frac{3^{2-n}n(1-n)(n+3)(2-n)^{n-3}}{4c_{1}H_{0}^{n-1}}$.\\
In the investigation of holographic dark energy density with the Hubble's horizon as a potential IR-cut-off, we presented the energy density, equation of state parameter as a function of red-shift for the combined dataset of $CC+SC+BAO$ in Figs. \ref{HD} and \ref{w}. As shown in Fig. \ref{HD}, the energy density is an increasing function of $z$ and behaves positively for model parameter values which are constrained by the combined datasets of $CC+SC+BAO$.\\ The EoS parameter is a crucial cosmological parameter for comprehending the nature of the Universe and its history across time, as it is stated in Eq. (\ref{e30}). The fluid's behaviour and its impact on the universe's expansion are determined by the value of the EoS parameter which are constrained by the combined datasets of $CC+SC+BAO$. For instance, a fluid is said to have non-relativistic matter and acts like dust if $\omega_{\Lambda}= 0$. The fluid, however, is referred to as relativistic matter and behaves like radiation if $\omega_{\Lambda}= 1/3$. If $\omega_{\Lambda}<1/3$, the fluid is assumed to have negative pressure and is in charge of the Universe's accelerated expansion, a phenomenon connected to DE which incorporates the quintessence $(-1<\omega_{\Lambda}<-1/3)$, cosmological constant $( \omega_{\Lambda}=-1$), and phantom era $(\omega_{\Lambda}<-1$). In this analysis, the model acts in the present like a quintessence at $z = 0$ and at $z = -1$ is $-1<\omega_{\Lambda}<-1/3$ which indicates an accelerating phase (see Fig. \ref{w}).
\noindent
Squared velocity of sound,
\begin{equation}\label{e31}
v_{s}^{2}=\frac{\alpha_{2}(3-n)}{2}\left(\frac{1}{1+z}\right)^{\frac{3(n-1)}{2-n}}
\end{equation}
Total density parameter,
\begin{equation}\label{e32}
\Omega=1+\frac{1}{3H_0^2}\left(\frac{1}{1+z}  \right)^{\frac{3(n-1)}{2-n}}
\end{equation}
The present examined model's squared sound velocity and total density parameter are given in Eqs. (\ref {e29}) and (\ref{e30}), respectively, and their behaviour is clearly shown in Figs. (\ref{V}) and (\ref{O}), respectively. Many authors \cite{28,51,52} have looked at the squared velocity of sound and total density parameter for this reason: the total density parameter confirms the model is either flat or open or closed respectively for $\Omega=1, 1$ and $>1$ (according to the observational results of SNe-Ia and Cosmic Microwave Background Anisotropy Spectrum) and the squared velocity of sound confirms the model is either stable or unstable (if the squared velocity of sound lies between zero and one, then the model is stable otherwise unstable). The Fig. \ref{V} depicts the progression of the squared sound velocity with respect to red-shift (z). With the cosmic expansion of the Universe, $v_{s}^{2} > 0$, it is evident that the model is stable whereas according to Eq. (\ref{e30}), the universe eventually becomes flat when the total density parameter approaches a constant value say one. The resulting model is well-suited with the observational results because our model predicts a flat Universe for a very long period of time and the current Universe is extremely near to becoming flat where $\Omega \approx 1$. Fig. \ref{O} depicts how the total density parameter changes across evolution.

\begin{figure}[H]
   \begin{minipage}{0.49\textwidth}
     \centering
   \includegraphics[scale=0.8]{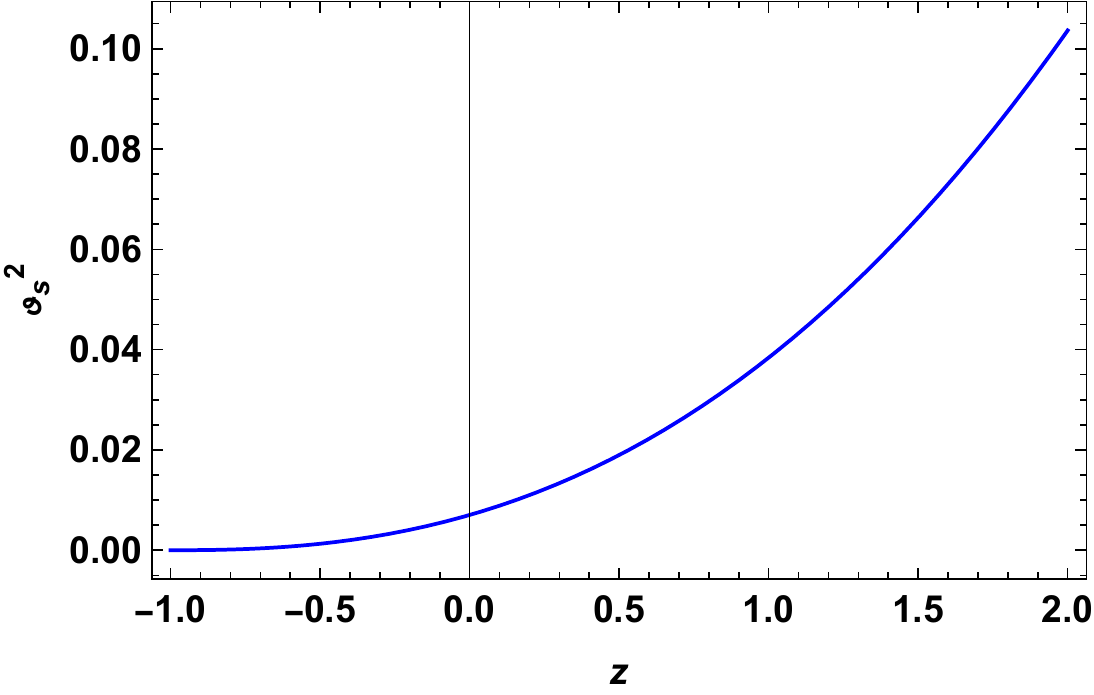}
\caption{The figure shows the evolution of $v_{s}^{2}(z)$ in Hubble's horizon as a potential IR Cut-off.}\label{V}
   \end{minipage}\hfill
   \begin{minipage}{0.49\textwidth}
     \centering
 \includegraphics[scale=0.84]{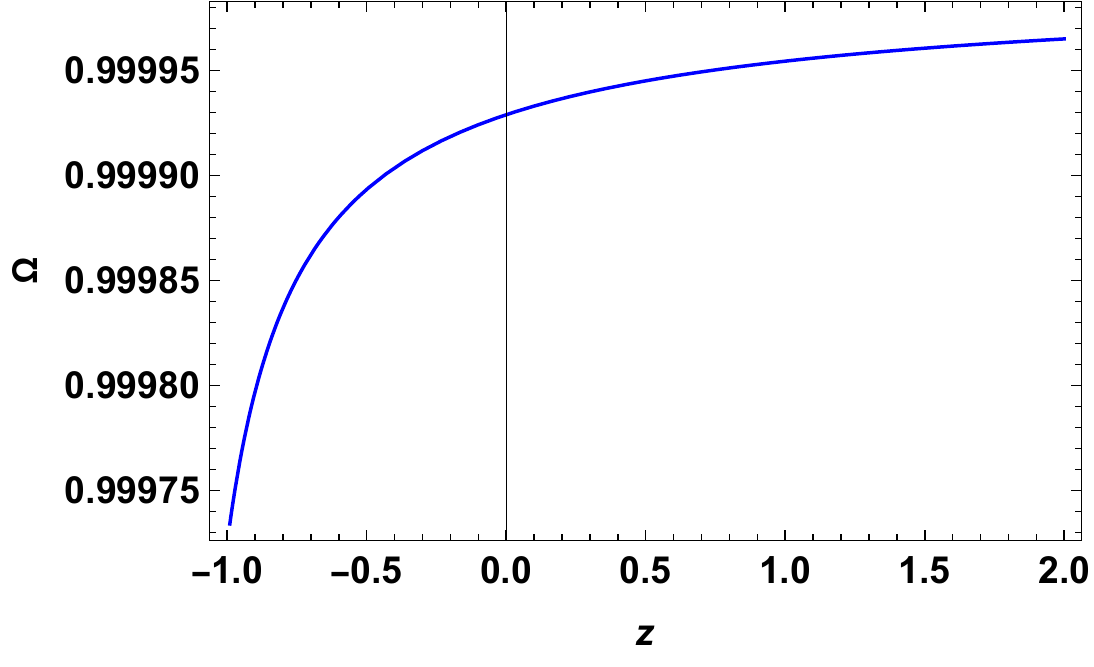}
\caption{The figure shows the evolution of $\Omega(z)$ in Hubble's horizon as a potential IR Cut-off.}\label{O}
   \end{minipage}
\end{figure}

\begin{figure}[H]
   \begin{minipage}{0.29\textwidth}
     \centering 
   \includegraphics[scale=0.5]{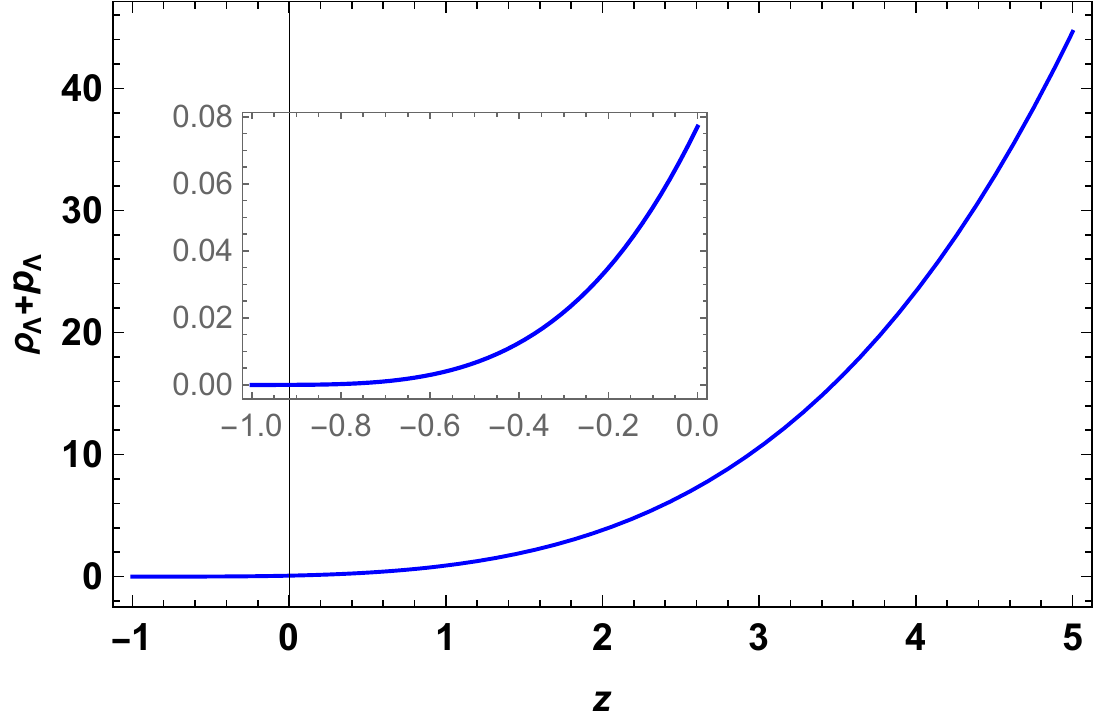}
 \centering ($a$)
   \end{minipage}\hfill
   \begin{minipage}{0.29\textwidth}
     \centering
 \includegraphics[scale=0.52]{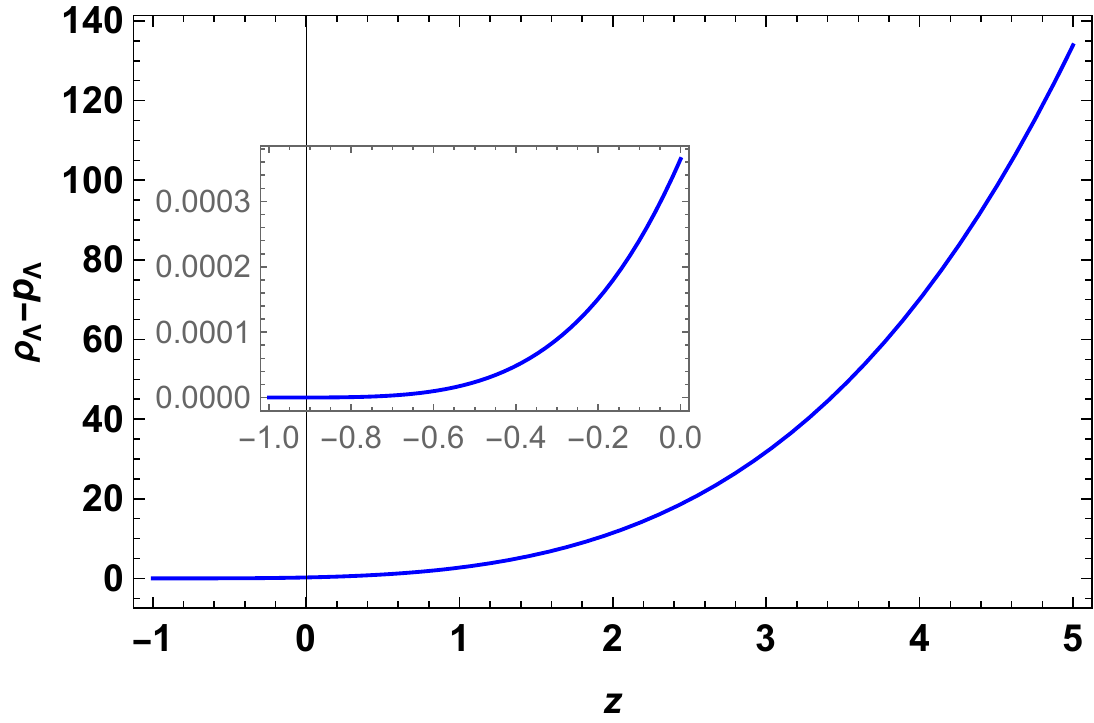}
 \centering ($b$)
   \end{minipage}\hfill
   \begin{minipage}{0.29\textwidth}
     \centering
 \includegraphics[scale=0.5]{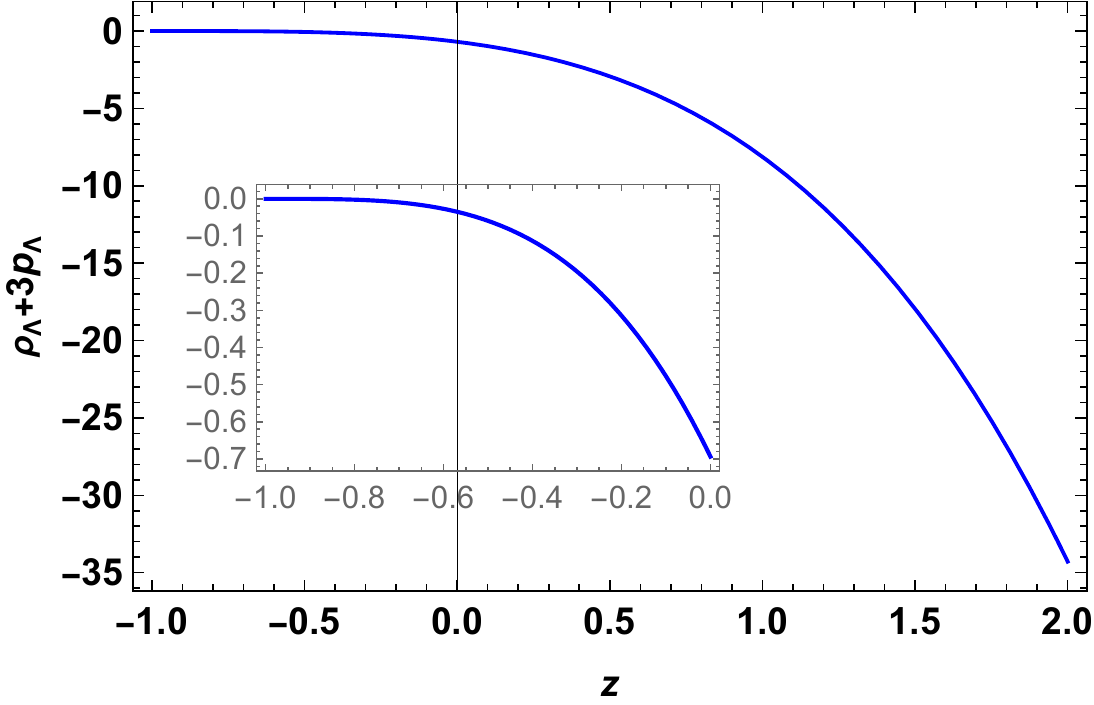}
 \centering ($c$)
   \end{minipage}
\caption{The figure shows the respective evolutions of $\rho_{\Lambda}(z)+p_{\Lambda}(z)$, $\rho_{\Lambda}(z)-p_{\Lambda}(z)$ and $\rho_{\Lambda}(z)+3p_{\Lambda}(z)$ in left panel $(a)$, middle  panel $(b)$ and right panel $(c)$ in Hubble's horizon as a potential IR Cut-off.}\label{EC}
\end{figure}

Energy conditions\\
Eqs. (\ref{e16}) and (\ref{e29}), when applied, yielded the following observations for the NEC, DEC, and Strong energy conditions as\\
NEC
\begin{equation}\label{e33}
\rho_{\Lambda}+p_{\Lambda}=3H^2 \left[1+\alpha_{2} \left(\frac{1}{1+z}\right)^{\frac{3(n-1)}{(2-n)}}\right]
\end{equation}
DEC
\begin{equation}\label{e34}
\rho_{\Lambda}-p_{\Lambda}=3H^2 \left[1-\alpha_{2} \left(\frac{1}{1+z}\right)^{\frac{3(n-1)}{(2-n)}}\right]
\end{equation}
SEC
\begin{equation}\label{e35}
\rho_{\Lambda}+3p_{\Lambda}=3H^2 \left[1+3\alpha_{2} \left(\frac{1}{1+z}\right)^{\frac{3(n-1)}{(2-n)}}\right]
\end{equation}
The accompanying Fig. \ref{EC} depicts the behaviour of the Null (left panel ($a$)), Dominant (middle panel ($b$)), and Strong (right panel ($c$)) energy conditions of the current cosmological model. The graphic showed that $\rho_{\Lambda}+p_{\Lambda}$, $\rho_{\Lambda}-p_{\Lambda}$, and $\rho_{\Lambda}+3p_{\Lambda}$ indicated that the SEC was violated whereas NEC and DEC were energy conditions that were confirmed.

\subsection{Granda-Oliveros cut-off}

\begin{figure}[H]
   \begin{minipage}{0.49\textwidth}
     \centering
   \includegraphics[scale=0.8]{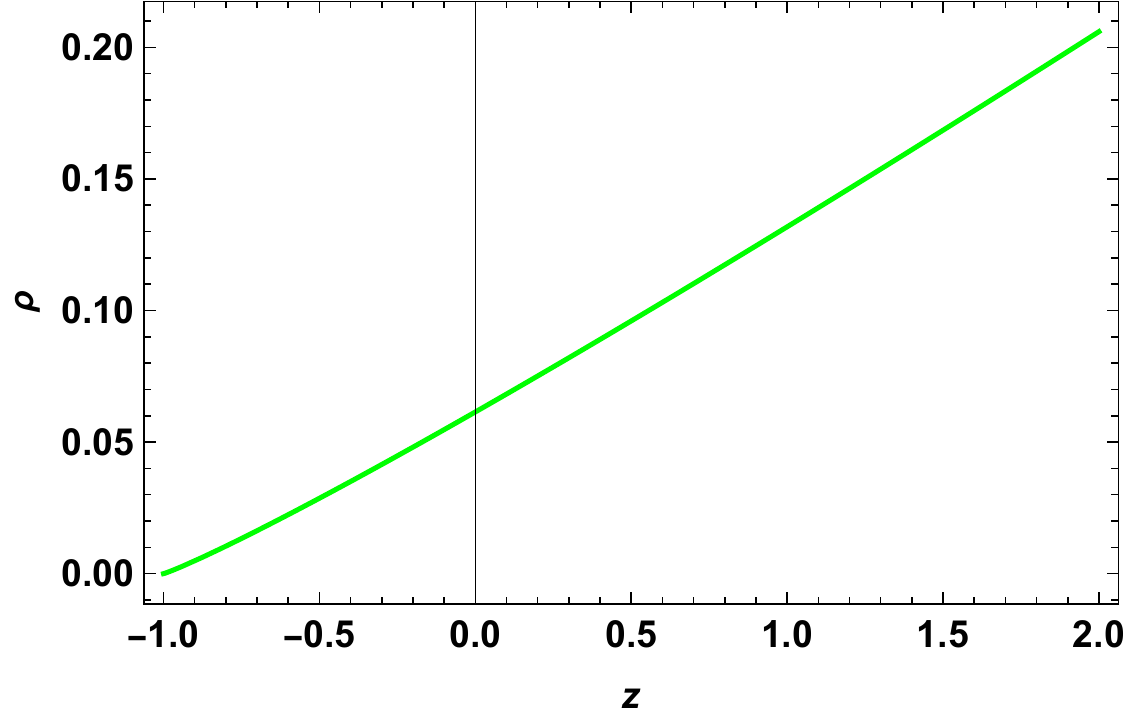}
\caption{The figure shows the evolution of $\rho(z)$ in Granda-Oliveros horizon as a potential IR cut-off.}\label{GOD}
   \end{minipage}\hfill
   \begin{minipage}{0.49\textwidth}
     \centering
 \includegraphics[scale=0.84]{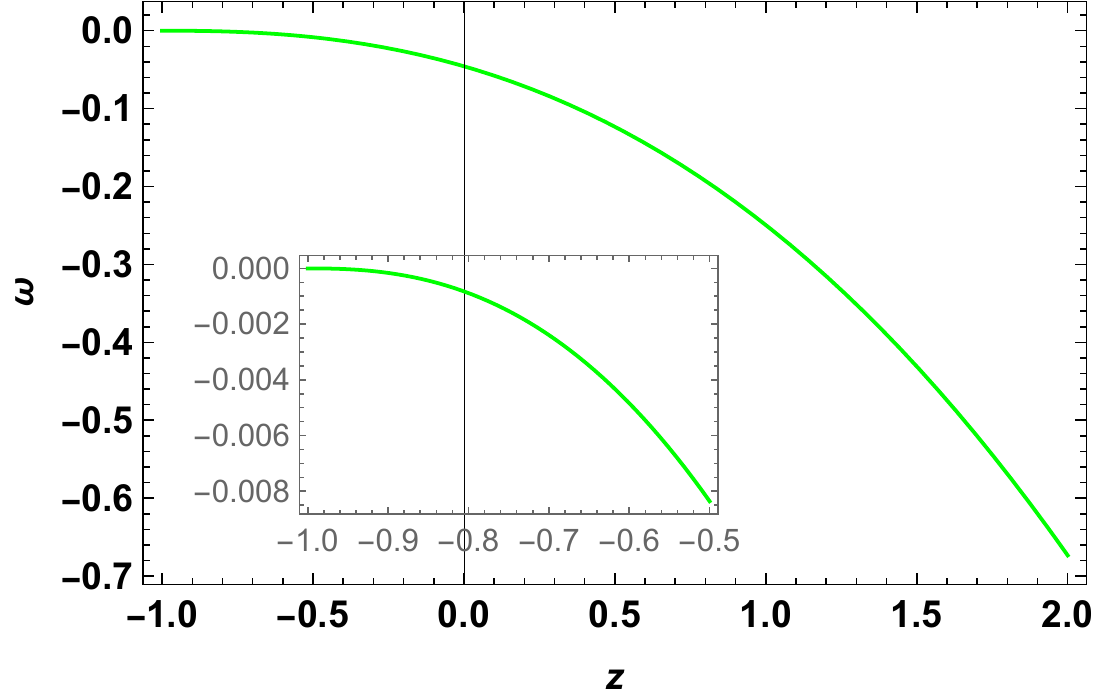}
\caption{The figure shows the evolution of $\omega(z)$ in Granda-Oliveros horizon as a potential IR cut-off.}\label{GOw}
   \end{minipage}
\end{figure}

Here, some cosmological parameter are determined by using Granda-Oliveros horizon ($L=\left( \alpha H^2+\beta \dot{H}\right)^{\frac{-1}{2}}$) as a potential IR-cut-off.\\
In view of Eq. (\ref{e15}),  the expression of holographic dark energy density from Eq. (\ref{e29}) is obtained as,
\begin{equation}\label{e36}
\rho_{\Lambda}= \alpha_{3} \left(1+z\right)^{\frac{6}{2-n}}
\end{equation}
where $\alpha_{3}=\frac{3H_0^2}{(2-n)}\left[\alpha (2-n)-3\beta\right]$.\\
EoS parameter,
\begin{equation}\label{e37}
\omega_{\Lambda}=\frac{\alpha_{2}}{(2-n)\left[\alpha (2-n)-3\beta\right]}\left(\frac{1}{1+z}\right)^{\frac{3(n-1)}{2-n}}
\end{equation}
With the combined dataset of $CC+SC+BAO$, we provided the energy density, equation of state parameter as a function of red-shift in  Figs. \ref{GOD} and  Figs. \ref{GOw}  in the examination of holographic dark energy density with the Granda-Oliveros horizon as a probable IR-cut-off. The energy density, as shown in Fig. \ref{GOD}, is an increasing function of $z$ and has a favourable behaviour for model parameter values restricted by the combined datasets of $CC+SC+BAO$. Eq. (\ref{e37}) states that the EoS parameter is a fundamental cosmological parameter for understanding the nature of the Universe and its history across time. The EoS parameter's value, which is constrained by the combined datasets of $CC+SC+BAO$, determines the behaviour of the fluid and its effect on the universe's expansion. As an illustration, our model behaves as a quintessence at $z = 0$ and is $-1<\omega_{\Lambda}<-1/3$ at $z = -1$, which denotes an accelerating phase (see Fig. \ref{GOw}).

Squared velocity of sound,
\begin{equation}\label{e38}
v_{s}^{2}=\frac{(3-n)\alpha_{2}}{2(2-n)\left[\alpha (2-n)-3\beta\right]}\left(\frac{1}{1+z}\right)^{\frac{3(n-1)}{2-n}}
\end{equation}
\begin{equation}\label{e39}
\Omega=\alpha-\frac{3\beta}{(2-n)}+\frac{c_2}{3H_{0}^{2}}
\left(\frac{1}{1+z}  \right)^{3\delta^{\prime \prime}}
\end{equation}
where $\delta^{\prime \prime}=\delta+\frac{n}{2-n}$.\\
The present examined model's squared sound velocity and total density parameter are given in Eqs. (\ref{e38}) and (\ref{e39}), respectively, and their behaviour is clearly shown in  Figs.\ref{V} and  Figs.\ref{O}, respectively. Fig.(\ref{V}) shows how the squared sound velocity changes over time in relation to redshift ($z$). According to Eq. (\ref{e39}), the cosmos eventually becomes flat when the total density parameter approaches a fixed value, say one. Yet, with the cosmic expansion of the universe, $v_s^2 > 0$, it is obvious that the model is stable. Our model predicts a flat Universe over a very long time, and the current Universe is very close to being flat where $\Omega \approx 1$. As a result, the derived model fits the observational data well.
\begin{figure}[H]
   \begin{minipage}{0.49\textwidth}
     \centering
   \includegraphics[scale=0.8]{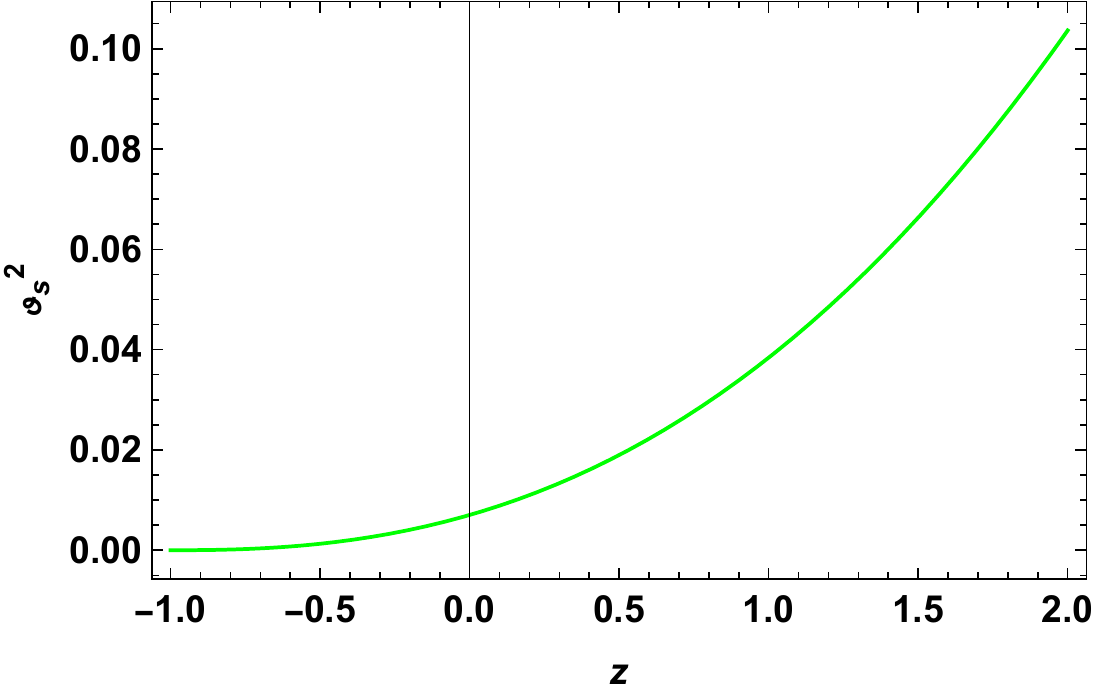}
\caption{The figure shows the evolution of $v_{s}^{2}(z)$ in Granda-Oliveros horizon as a potential IR cut-off.}\label{V}
   \end{minipage}\hfill
   \begin{minipage}{0.49\textwidth}
     \centering
 \includegraphics[scale=0.84]{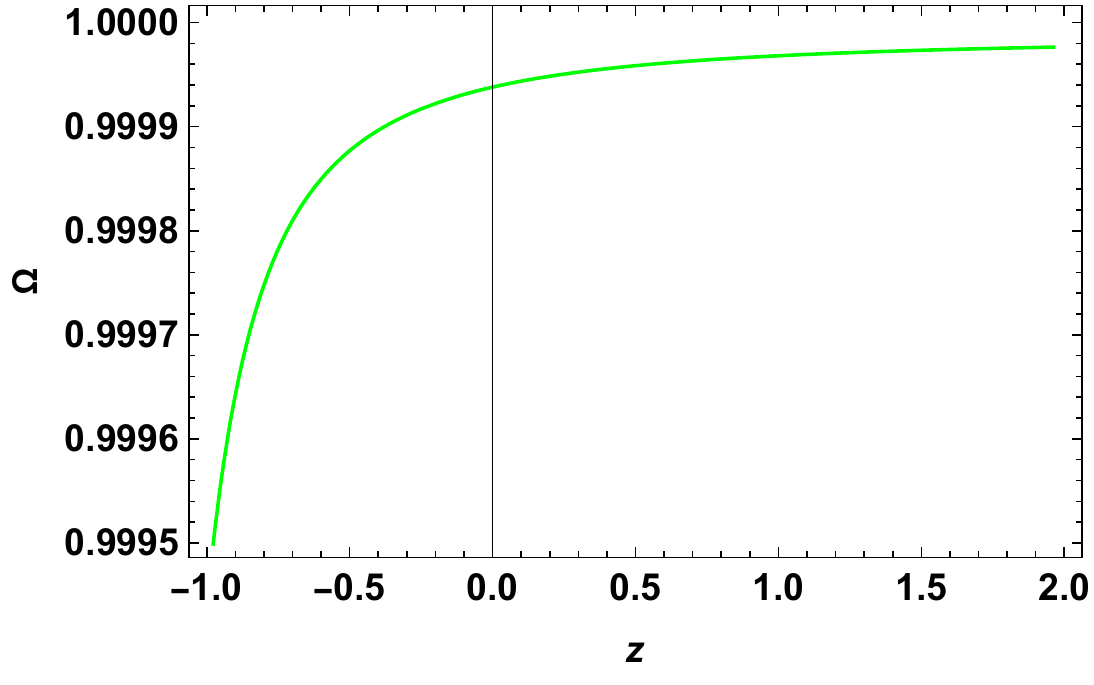}
\caption{The figure shows the evolution of $\Omega(z)$ in Granda-Oliveros horizon as a potential IR cut-off.}\label{O}
   \end{minipage}
\end{figure}

Energy conditions\\
Eqs. (\ref{e16}) and (\ref{e36}), when applied, yielded the following observations for the NEC, DEC, and Strong energy conditions as\\
NEC
\begin{equation}\label{e40}
\rho_{\Lambda}+p_{\Lambda}=3H^2 \left[\alpha-\frac{3\beta}{2-n}+\alpha_{2} \left(\frac{1}{1+z}\right)^{\frac{3(n-1)}{(2-n)}}\right]
\end{equation}
DEC
\begin{equation}\label{e41}
\rho_{\Lambda}-p_{\Lambda}=3H^2 \left[\alpha-\frac{3\beta}{2-n}-\alpha_{2} \left(\frac{1}{1+z}\right)^{\frac{3(n-1)}{(2-n)}}\right]
\end{equation}
SEC
\begin{equation}\label{e42}
\rho_{\Lambda}+3p_{\Lambda}=3H^2 \left[\alpha-\frac{3\beta}{2-n}+3\alpha_{2} \left(\frac{1}{1+z}\right)^{\frac{3(n-1)}{(2-n)}}\right]
\end{equation}

The behaviour of the present cosmological model's Null (left panel ($d$), Dominant (middle panel ($e$), and Strong (right panel ($f$)) energy conditions is shown in the following Fig. \ref{EC1}. The graph demonstrated that the SEC was violated, but the NEC and DEC were energy conditions that were confirmed.

\begin{figure}[H]
   \begin{minipage}{0.29\textwidth}
     \centering 
   \includegraphics[scale=0.5]{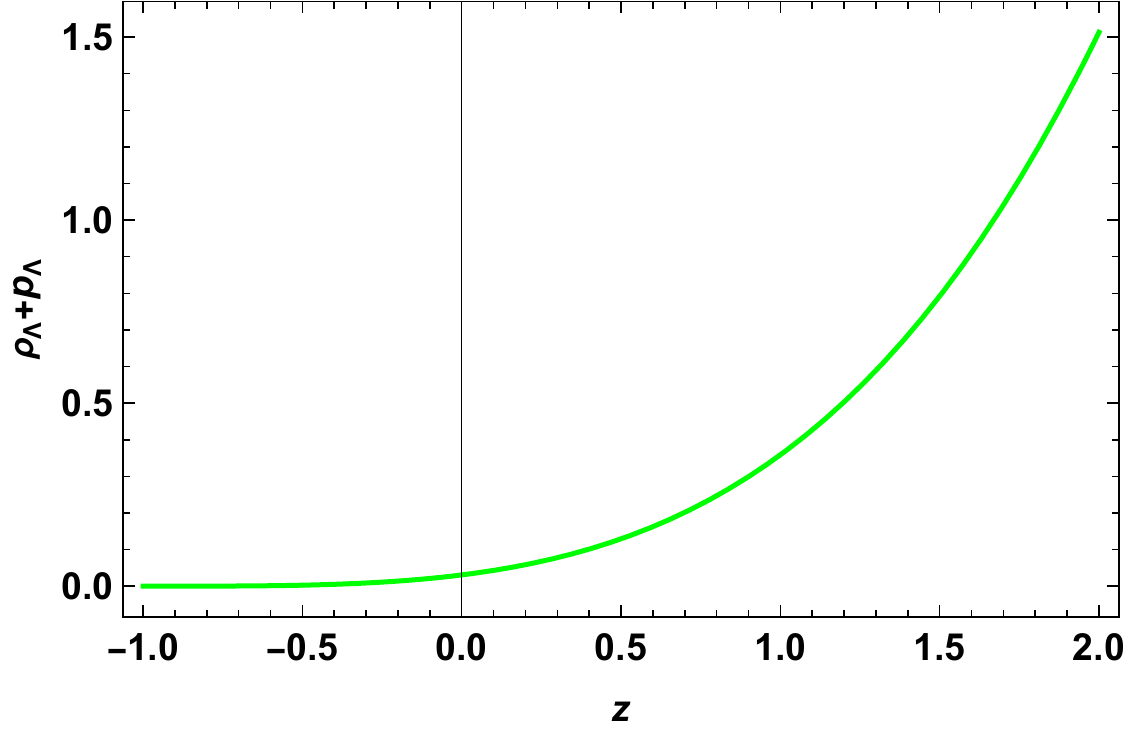}
 \centering ($d$)
   \end{minipage}\hfill
   \begin{minipage}{0.29\textwidth}
     \centering
 \includegraphics[scale=0.52]{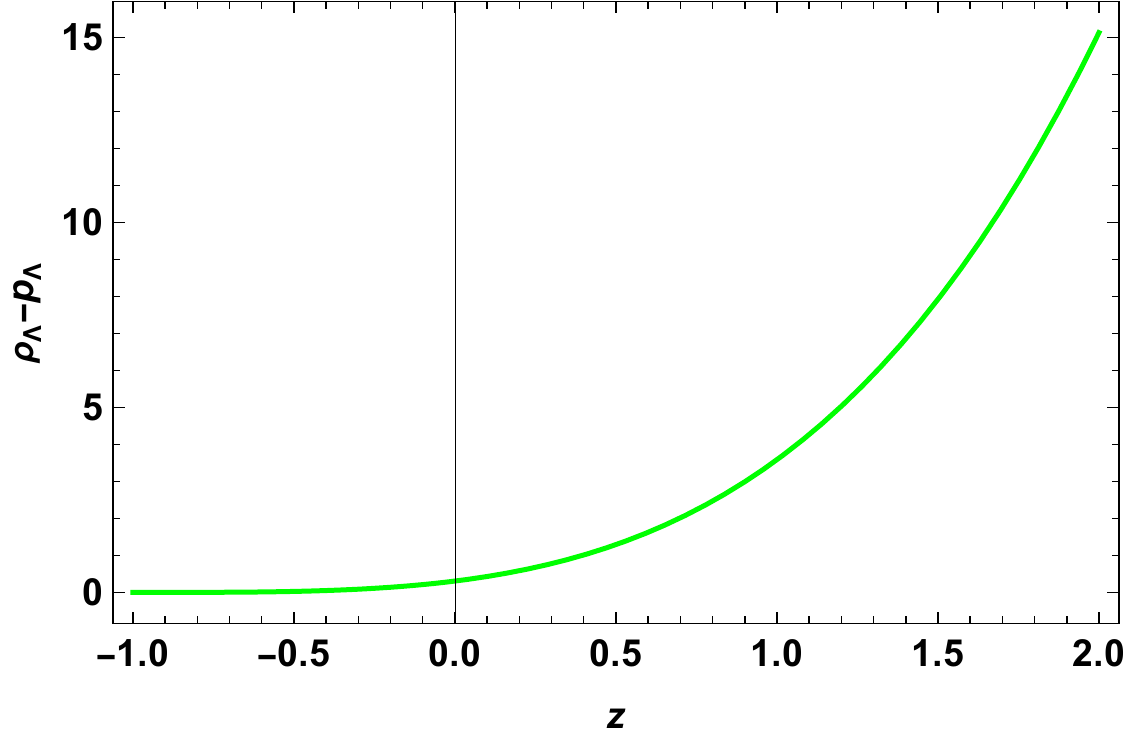}
 \centering ($e$)
   \end{minipage}\hfill
   \begin{minipage}{0.29\textwidth}
     \centering
 \includegraphics[scale=0.5]{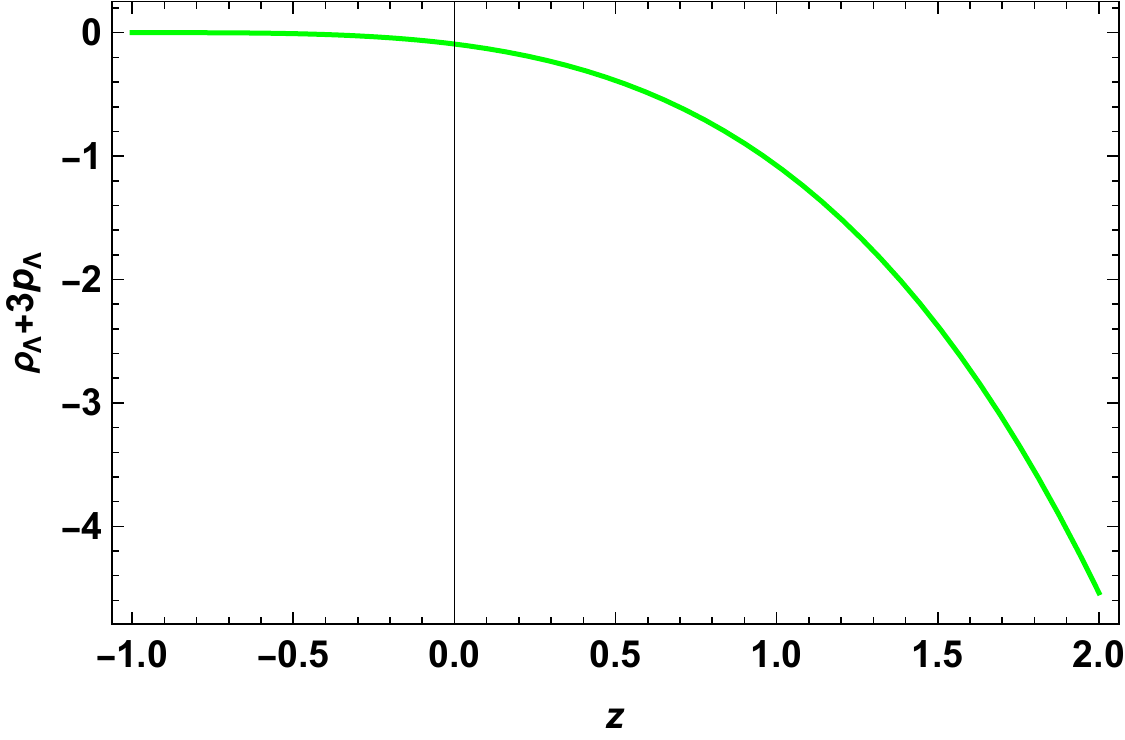}
 \centering ($f$)
   \end{minipage}
\caption{The figure shows the respective evolutions of $\rho_{\Lambda}(z)+p_{\Lambda}(z)$, $\rho_{\Lambda}(z)-p_{\Lambda}(z)$ and $\rho_{\Lambda}(z)+3p_{\Lambda}(z)$ in left panel $(d)$, middle  panel $(e)$ and right panel $(f)$ in  Granda-Oliveros horizon as a potential IR cut-off.}\label{EC1}
\end{figure}

\section{Conclusion}
The present study investigates observational limitations on the holographic dark energy model as a viable candidate for the Hubble and Granda-Olivero IR cut-offs. In this model we have consider the  non-static plane-symmetric space-time, and provide a comprehensive analysis of several observables, including energy density, equation of state parameters, and energy conditions. In this model we  have been expressed several important cosmological parameters in terms of redshift '$z$' and shown graphically for the understanding of the universe's expansion and development. The resultant model is also compared to the $\Lambda$CDM model. In our derived model we have studied the observational limits of the holographic dark energy with the Ricci cutoff, has been determined  from the combined CC+SC+BAO data.

We plotted the 1$\sigma$ and 2$\sigma$ confidence levels, as well as the one-dimensional two-dimensional probability distribution.In the continuation of our study, we displayed the $H(z)-z$ figure, which compares the theoretical function of H(z) with 31 observations of the Hubble parameter $H(z)$. This graph shows the appropriate agreement between our models and the observed data.
 We have summarized our results as follows:
\begin{itemize}
	\item
The plots in Fig.1 and Fig.2,$ H(z)-z$ and $\mu(z)-z$ show the fitting of our model and compared with the $\Lambda$CDM model together with the error bars for the CC+SC+ BAO datasets.
\item
 The plots of Fig.3  showns  the 2-D contour plot and showing best fit values $H_{0}=70.767$ and  $n=-3.036$ by obtained from emcee codes for the CC+SC+BAO datasets with 1-$\sigma$ and 2-$\sigma$ errors.
\item
The energy density of HDE $\rho$ is positive increasing function with redshift $z$. We found that this behaviour is similar in both IR cut-offs (Hubble and Granda Olivor) for combined data CC+SC+BAO as seen in Fig.4$\&$ 9 respectively.
\item

According to Fig. 5 $\&$ 10, the trajectories of the EoS parameter demonstrate how the vacuum era of the cosmos evolved during the transition from the phantom zone to the quintessence region.
The figures show how the Equation of State parameter $\omega$ changes with redshift. EoS parameter has been found to be around  $\omega=-0.75$ and $\omega=-0.87$ at the present time in both IR cut-offs (Hubble and Granda Olivor) for combined data CC+SC+BAO, which is consistent with Riess \cite{1} and indicates that $\Omega$ is now in the quintessence area.

\item
Fig.6,7 $\&$ 11,12 depicts the behaviour of velosity of sound. The stability condition happens when the positive squared sound speed $v_s^{2}>0$. Our derived model is stable for both IR cut off. Similarly  we have also notice that the density parameter $\Omega$ is positive throughout the evolution in both IR cut-offs as seen in Fig.7, 11 for  combined data CC+SC+BAO .
\item
We have discussed some physical properties of the model as well as the evolution of physical parameters and Energy Conditions. It is observed that NEC and DEC do not violate, whereas SEC violate , which produces a repulsive force. In Fig.8$\&$ 13 the violation of SEC demonstrates the viability of our model, as stated in \cite{53}. ECs have provided us with unique insights into the
  deep structure for space and time in the cosmic spacetime evolution processes. In the current model, NEC and DEC are validated, but SEC is violated due to the requirement of cosmic acceleration in both cutoffs. \\

 Our study and the accumulated observational data are compatible with the possibility of the holographic dark energy the theory with Hubble and GO cut-off.

  \end{itemize}

\end{document}